\documentclass[12pt,fleqn]{article}

\usepackage{latexsym}
\usepackage{amsmath}
\usepackage{amssymb}

\renewcommand{\[}{$$}
\renewcommand{\]}{$$%
\par  \noindent  \hspace{-0.4em}}

\newcommand{\rf}[1]{(\ref{#1})}

\newcommand{\ba}{\begin{array}}
\newcommand{\ea}{\end{array}}

\newcommand{\be}{\begin{equation}}
\newcommand{\ee}{\end{equation}}

\newcommand{\ods}{\par \vspace{0.5cm} \par}
\newcommand{\no}{\noindent}

\newcommand{\const}{{\rm const}}

\newcommand{\R}{{\mathbb R}}

\newcommand{\dis}{\displaystyle}

{\normalsize \hfill {\mbox{$\Box$}} \par  \vspace{1.5ex}}

\newtheorem{prop}{Proposition}[section]

\newtheorem{cor}[prop]{Corollary}

\newtheorem{Ex}[prop]{Example}

\begin{document}

\title{\bf A direct approach to the construction of standard
and non-standard Lagrangians for dissipative dynamical systems with variable coefficients
}
\author{Jan L.\ Cie\'sli\'nski\thanks{e-mail:
\tt janek\,@\,alpha.uwb.edu.pl}
\\ Tomasz Nikiciuk\thanks{e-mail:
\tt niki\,@\,alpha.uwb.edu.pl}
\\ {\footnotesize Uniwersytet w Bia\l ymstoku,
Wydzia\l \ Fizyki,}
\\ {\footnotesize ul.\ Lipowa 41, 15-424 Bia\l ystok, Poland}
}

\date{}

\maketitle

\begin{abstract}
We present a direct approach to the construction of Lagrangians for a large class of one-dimensional dynamical systems with a simple dependence (monomial or polynomial) on the velocity. We rederive and generalize some recent results and find Lagrangian formulations which seem to be new. Some of the considered systems (e.g., motions with the friction proportional to the velocity and to the square of the velocity) admit infinite families of different Lagrangian formulations.  
\end{abstract}

   \par


\ods

\section{Introduction}

In recent papers \cite{MRS,Mus} a problem of finding Lagrangian description for a large class of one-dimensional
dissipative (or dissiptive-looking) systems was discusssed. The discussion was far from being exhaustive. In this paper
we present a different, more direct, approach to the problem of the construction of Lagrangians for
disipative (or dissipative-looking) systems. We simply assume some general form of the Lagrangian and then check
the resulting Euler-Lagrange equations.

The inverse problem of Lagrangian mechanics is concerned with the question whether
a given system of second-order ordinary differential
equations $\ddot q^i = f^i (t, q, \dot q)$
can be derived from a variational principle \cite{Lop}. In other words, one tries to find a Lagrangian for this system. This problem was studied in XIX century by Helmholtz (see \cite{Sar}) and by Darboux who proved that in the one-dimensional case the Lagrangian always exists \cite{Dar}. The inverse problem in two-dimensional case was solved by Douglas \cite{Dou}, while the general case has been completed recently \cite{SCM}.

Dissipative systems were long believed to be `beyond variational treatment' \cite{Lan}, which is to some extent true if we insist on physical interpretation of the Hamiltonian and canonical momenta, compare \cite{Bau}. However, by relaxing these requirements one can
get variational interpretation of numerous dissipative system \cite{Bat,Ri1,Ri2,DY}.

In this paper we focus on more elementary issues, namely on providing explicit Lagrangian description for a large class of one-dimensional differential equations of second order with a simple (e.g., polynomial) dependence on the velocity $\dot x$.

\section{Standard Lagrangians}

Standard Lagragians (known also as `natural' or `of mechanical type') are  quadratic forms with respect to $\dot x$ (the dot denotes the differentiation with respect to $t$).
In the one-dimensional case we can easily obtain all equations of motions
corresponding to standard Lagrangians.  We assume
\be  \label{stand-L}
 {\cal L} = \frac{1}{2} P (x, t) {\dot x}^2 + Q (x, t) {\dot x} + R (x,t) \ ,
\ee
The Euler-Lagrange equations yield
\be
{\ddot x} + \frac{P_x}{2 P} {\dot x}^2 + \frac{P_t}{P} {\dot x} + \frac{Q_t - R_x}{P} = 0 \ ,
\ee
where subscripts $x, t$ denote partial derivatives. 
As a consequence we immediately obtain the following proposition.

\ods
\begin{prop}  \label{prop-abc}
The equation of motion
\be  \label{stand-ruch}
\ddot x + a (x,t) {\dot x}^2 + b (x,t) \dot x + c (x,t) = 0
\ee
admits a Lagrangian description with a standard Lagrangian \rf{stand-L} iff
\be
b_x = 2 a_t \ .
\ee
Then \ $ P = \exp (2 \int^x a (\xi,t) d \xi ) $, and
\be  \label{quad-R}
 R = \int^x \left( Q_t (\xi, t) - c (\xi, t)  P (\xi, t) \right) d \xi \ ,
\ee
where $Q = Q(x,t)$ is an arbitrary function.

\end{prop}

All systems described by \rf{stand-L} have also the Hamiltonian description.
Indeed, computing generalized momentum
\be
  p = P \dot x + Q \ , \quad  \dot x = \frac{p - Q}{P}
\ee
we easily get the standard Hamiltonian $H = p \dot x - \cal L$:
\be
  H (x, p, t) = \frac{( p - Q (x, t))^2}{ 2 P (x, t) } - R (x, t) \ .
\ee
\ods

\begin{cor}  \label{cor-abc}

Special cases of Proposition~\ref{prop-abc}:

\begin{enumerate}

\item $P=P(t)$ and $Q \equiv 0$:
\[
\dis \ddot x + b (t) {\dot x} + c (x, t) = 0 \quad \Longrightarrow \quad
{\cal L} = \left( \frac{1}{2} {\dot x}^2 - \int^x c (\xi, t) d \xi \right) e^{\int^t b(\tau)d\tau}
\]
This is a generalization of Proposition 1 from \cite{Mus}. In the case of linear equations (i.e., $c = x \tilde c (t)$) we have:
\be   \label{ABH}
    {\cal L} = \left( \frac{1}{2} {\dot x}^2 - \frac{1}{2} {\tilde c} (t) x^2 \right) e^{\int^t b(\tau)d\tau} \ .
 \ee
In particular, we rederive the well known result \cite{Lem1,Lem2} for the damped  harmonic oscillator:
\[
 \dis \ddot x +  \gamma \dot x + \omega_0^2 x = 0 \quad \Longrightarrow \quad
{\cal L} = \frac{1}{2} e^{\gamma t} \left( {\dot x}^2 - \omega_0^2 x^2 \right) \ .
\]

\item $P = P (x) $  and $R \equiv 0$:
\[
\ddot x + a (x) {\dot x}^2 + c (x, t) = 0 \quad \Longrightarrow \quad
{\cal L} = \left( \frac{1}{2} {\dot x}^2 + {\dot x} \int^t c (x,\tau) d\tau \right)  e^{2 \int^x  a (\xi) d\xi}
\]

This formula simplifies for $c = c (x)$ (the case considered in \cite{MRS,Mus}). Then
\be
{\cal L} = \left( \frac{1}{2} {\dot x}^2 + t {\dot x} c (x) \right)  e^{2 \int^x  a (\xi) d\xi}
\ee
\item $P = P (x) $  and $Q \equiv 0$:
\[
\ddot x + a (x) {\dot x}^2 + c (x, t) = 0 \quad \Longrightarrow \quad
{\cal L} =  \frac{1}{2} {\dot x}^2 e^{2 \int^x  a (\xi) d\xi} -  \int^x c (\xi, t) e^{2 \int^\xi  a (y) d y} d\xi
\]
This is a generalization of the main result of \cite{MRS} and Proposition 3 from \cite{Mus}, where $c = c (x)$. Thus these results are extended on $t$-dependent function $c = c (x,t)$.

\item  $P = A (x) B (t)$:
\[
\ddot x + a (x) {\dot x}^2 + b (t) {\dot x} + c (x, t) = 0 \quad \Longrightarrow \quad
\cal L \ \ {\it is\ given\ by\ \rf{stand-L},\ where}
\]
$P = A B$, $A = \exp ( 2 \int^x a (\xi) d\xi)$, $B = \exp ( \int^t b(\tau)  d \tau )$, $R$ is given by \rf{quad-R},
and $Q$ is arbitrary.

\end{enumerate}
\end{cor}

\ods

\begin{Ex}[A particle accreting mass in a potential field]

We proceed to physical aspects of the equation
\be  \label{eq-accr}
\ddot x + b (t) {\dot x} + c (x, t) = 0 \ .
\ee
Following \cite{KRS}, where the damped harmonic oscillator is interpreted as harmonic oscillator with time-dependent mass, we define
\be
   m (t) = e^{\int^t b (\tau) d \tau } \ , \quad {\rm i.e.}, \quad b (t) = \frac{\dot m}{m} \ .
\ee
Then,
\be
   {\cal L} = \frac{1}{2} m (t) {\dot x}^2 - m (t) V (x,t) \ , \quad
 H = \frac{p^2}{2 m (t) } + m (t) V (x,t) \ ,
\ee
where $V (x,t) = \int^x c (\xi, t) d \xi$. Therefore, the equation \rf{eq-accr} can be considered either as a dissipative system or a particle with a prescribed mass time-dependence in an arbitrary potential (possibly time dependent).
\end{Ex}

\section{Reciprocal Lagrangians}

Reciprocal Lagrangians (i.e., inverses of standard-like Lagrangians)
were introduced and studied recently \cite{Mus,CRS,CSL}. If
\be
      {\cal L} = \frac{1}{L} \ , \quad L = L (x, \dot x, t) ) \ ,
\ee
then
\be  \label{Lrec}
{\ddot x} = \frac { \displaystyle 2 {\dot x} \frac{\partial L}{\partial \dot x} \frac{\partial L}{\partial x} -
{\dot x} L \frac{\partial^2 L}{\partial {\dot x} \partial x } + 2 \frac{\partial L}{\partial t}
\frac{\partial L}{\partial \dot x} - L \frac{\partial^2 L}{\partial t \partial \dot x} +
L \frac{\partial L}{\partial x}  }{ \displaystyle L \frac{\partial^2 L}{\partial {\dot x}^2 } - 2 \left( \frac{\partial L}{\partial \dot x}
 \right)^2   }
\ee
We confine ourselves to $L$ of the form
\be  \label{FnG}
L = F (x,t) {\dot x}^\nu + G (x,t) \ .
\ee
Substituting \rf{FnG} into \rf{Lrec} we obtain:
\be   \label{rruch2}
{\ddot
x}=\frac{p \dot{x}^{2\nu}+ q\dot{x}^{2\nu-1}+ r \dot{x}^{\nu}+ s\dot{x}^{\nu-1}+ w}{g\dot{x}^{\nu-2}-h\dot{x}^{2 \nu- 2}} \ ,
\ee
where
\[ \ba{l}
p:=(1+\nu)FF_{x}, \quad
q:=\nu FF_{t}, \quad
r:=(1+2\nu)FG_{x}+(1-\nu)F_{x}G,  \\[2ex]
s:=2\nu G_{t}F-\nu GF_{t}, \quad
w:=GG_{x} , \quad
g:=\nu(\nu-1)FG, \quad
h:=\nu(\nu+1)F^2 .
\ea \]
The case $\nu =1$, $F =1$ is discussed in \cite{CRS}, with  a special stress on $G$ quadratic in $x$ (leading to second-order Riccati equations), see also \cite{Mus}.
In the case $\nu = 1$ the equation \rf{rruch2} reduces to a special case of \rf{stand-ruch}:
\be  \label{rruch-FG}
{\ddot x}=- \frac{F_{x}}{F} \dot x^2- \frac{(F_{t}+3G_{x})}{2F} \dot x- \frac{(2G_{t}F-GF_{t}+GG_{x})}{2F^2}.
\ee

First, we confine ourselves to $t$-independent $F, G$. Then the coefficients $a, b, c$ by powers of $\dot x$ depends on $x$ only. They are not independent. Indeed,
\be  \label{abc}
   a = \frac{F'}{F} \ , \quad b = \frac{3G'}{2F} \ , \quad
c = \frac{GG'}{2F^2}  \ ,
\ee
where the prime denotes the differentiation with respect to $x$. 
Hence, substituting $G = 3 c F / b$ and $F' = a F$ to the last equation of \rf{abc}, we get a constraint on $a, b, c$, see \rf{2/9}.
\ods

\begin{prop}  \label{prop-rec1}
The equation \rf{stand-ruch} admits the Lagrangian description with ${\cal L} = (\dot x  F (x) + G (x) )^{-1}$ iff
\be  \label{2/9}
  c,_x + \left( a - \frac{b,_x}{b} \right) c = \frac{2}{9} b^2  \ .
\ee
Then, $F (x) = \exp (\int^x a (\xi) d \xi)$ and $G (x) = 3 c (x) F (x) / b (x)$.
\end{prop}

Therefore, we can choose arbitrary functions $a (x), b (x)$ and then $c$ have to satisfy the equation \rf{2/9}. Solving this equation we get:
\be
  c (x) = \frac{2}{9} b (x)
 \int^x b (\xi) \exp \left( \int_x^\xi a (y) d y \right) d \xi \ .
\ee
\ods
\begin{Ex}
Taking $a = 0$ and $b (x) = k x$ ($k = \const$) we obtain
\be
  c (x) = \frac{2}{9} k x \left( \frac{1}{2} k x^2 + \lambda \right) =
\frac{k^2 x^3 }{9} + \lambda_1 x \ ,
\ee
where $\lambda = \const$ and $\lambda_1 := \frac{2}{9} k \lambda$. This case corresponds exactly to a Li\'enard-type nonlinear oscillator which shows very unusual properties, like isochronous oscillations for $\lambda_1 > 0$ \cite{CSL}.
\end{Ex}
Another possibility is to choose arbitrary functions $b (x), c(x)$ and then $a (x)$ is given by:
\be
 a = \frac{b,_x}{b} - \frac{c,_x}{c} + \frac{2 b^2}{9 c} \ .
\ee
Proposition~\ref{prop-rec1} generalizes Propositions 4 and 5 from \cite{Mus}.
The case $\nu = 1$ contains other interesting subcases. Indeed, assuming
$ F = f (t)$, $G = x g (t)$
the equation \rf{rruch-FG} can be reduced to the linear equation:
\be  \label{rruch-mmf}
  \ddot x + b (t) \dot x + c(t) x = 0 \ ,
\ee
where
\be  \label{bcfg}
  b = \frac{\dot f + g}{2 f} \ , \qquad c = \frac{2 f \dot g - g \dot f + g^2}{2 f^2} \ .
\ee
The system \rf{bcfg} expresses $b, c$ in terms of $f, g$. It turns out that these equations can be inverted. Given $b, c$ we may compute corresponding $f, g$. Indeed, substituting  $g = 2 f b - \dot f $ into the second equation we get the inhomogeneous linear equation:
\be
   \frac{d }{d t} \left( \frac{\dot f}{f} \right) + b  \
\frac{\dot f}{f} = 2 \dot b - c \ ,
\ee
which can be solved in quadratures in a standard way:
\be  \label{quad-f}
   f (t) = \exp \left(  \int^t \left( \int^z \left( 2 \dot b (\tau) - c (\tau) \right)
\left( \exp \int_z^\tau b (y) d y \right) d \tau \right) d z \right)  .
\ee
\ods

\begin{prop}
The equation \rf{rruch-mmf} (for any $b (t), c(t)$) admits a Lagran\-gian description with the reciprocal Lagrangian of the form ${\cal L} = (\dot x  f (t) + x g (t) )^{-1}$, where $f$ is given by \rf{quad-f} and $g = 2 f b - \dot f$.
\end{prop}

Therefore, any equation of the form \rf{rruch-mmf} (including equations of mathematical physics, like Airy, Bessell, Hermite or Legendre equation) admits at least two different Lagrangians: standard one (see \rf{ABH}) and reciprocal.
\ods
\begin{Ex}
We get another simple case taking $F (t) = f_0 e^{2kt}$ and $G (t) = g_0 e^{kt}$. Then \rf{rruch-FG} reduces to $\ddot x + k \dot x = 0$.
\end{Ex}

\ods
In the case $\nu =2$ the Lagrangian \rf{FnG} yields more complicated equation:
\be
{\ddot x}=\frac{3FF_{x}\dot x^4+2FF_{t}\dot x^3 +(5FG_{x}-F_{x}G)\dot x^2+(4G_{t}F-2GF_{t}) \dot x + GG_{x} }{2F(G-3F{\dot x}^2)}
\ee
In the particular case $G = G (t)$, $F = f (x) G^3$ we obtain:
\be
 \ddot x = \frac{f'}{2 f} {\dot x}^2 + \frac{\dot G}{G} \dot x \ .
\ee

\ods
\begin{prop}
The equation
\be  \label{kwadlin}
 \ddot x + a (x) {\dot x}^2 + b (t) \dot x = 0
\ee
admits a Lagrangian description with a Lagrangian given by $( F {\dot x}^2 + G)^{-1}$, where
$G (t) = \exp ( - \int^t b (\tau) d \tau )$, $F (x, t) = \exp ( - 3 \int^t b (\tau) d \tau - 2 \int^x  a (\xi) d\xi)$.
\end{prop}

Therefore, the equation \rf{kwadlin} admits at least two different Lagrangian descriptions: standard (compare the case 4 of Corollary~\ref{cor-abc}) and reciprocal.

\section{Lagrangians with a modified kinetic term}

In this section we consider generalizations of standard Lagrangians, where
the kinetic term ${\dot x}^2$ is replaced by some more general expression (and the term linear in $\dot x$ is absent). 

\subsection{Monomial case}

First, we assume the monomial case:
\be  \label{lagr-FG}
 {\cal L} = F (x, t) {\dot x}^\mu - G (x, t) .
\ee
The equation of motion reads
\be \label{ruch-FGmu}
 \ddot x = - \frac{{\dot x}^2  F,_x }{\mu F} - \frac{\dot x  F,_t}{(\mu-1) F} -  \frac{{\dot x}^{2-\mu}  G,_x}{\mu (\mu-1) F} \ .
\ee
\ods
\begin{prop}  \label{prop-abcmu}
The equation of motion
\be  \label{ruch-abc}
\ddot x + a (x,t) {\dot x}^2 + b (x,t) \dot x + c (x,t) {\dot x}^{2-\mu} = 0
\qquad (\mu \neq 0, 1)
\ee
admits a Lagrangian description with the Lagrangian \rf{lagr-FG} iff
\be
(\mu-1) b,_x = \mu a,_t \ .
\ee
Then \ $ F = \exp (\mu \int^x a (\xi,t) d \xi ) $ \ and \
$ G = \mu (\mu - 1) \int^x  c (\xi, t)  F (\xi, t)  d \xi$.
\end{prop}

The proof follows directly by comparing \rf{ruch-abc} with \rf{ruch-FGmu}. Another result is obtained by assuming $F = F(x)$ and $G = G (x)$.

\ods
\begin{prop}
The equation
\be
\ddot x = - a (x) {\dot x}^2 - c (x) {\dot x}^\nu      \qquad  (\nu \neq 1, 2)
\ee
admits for any $a (x), c (x)$ a Lagrangian description. The Lagrangian  reads
\be
{\cal L} = F (x) {\dot x}^{2 - \nu} - G (x) \ ,
\ee
where
\[
F (x) = \exp \left( (2-\nu) \int^x a (\xi) d \xi \right)  , \quad
G (x) = (2 - \nu)(1 - \nu) \int^x c (\xi) F (\xi) d \xi   .
\]
\end{prop}
\ods
\begin{cor}  \label{cor-n-param}
Taking $c (x) = 0$, $a (x) = k = \const$, and denoting $n = \nu - 2$,  we obtain (for $n \neq 0$)
\[
 \ddot x + k {\dot x}^2 = 0 \quad \Longrightarrow \quad {\cal L} = C {\dot x}^n e^{n k x} \ .
\]
\end{cor}

\subsection{General case}

Let us  consider a class of standard-like Lagrangians with
quadratic kinetic terms replaced by an arbitrary smooth function of $\dot x$.
\be
 {\cal L} = F (x, t) \psi (\dot x) + G (x, t) \ .
\ee
The equation of motion reads
\be
{\ddot x} + \frac{( F_t + {\dot x} F_x ) \psi'  - F_x \psi - G_x}{F \psi''} = 0 \ .
\ee
Assuming  $F =F_0=\const$ we get the equation
\be
{\ddot x} = \frac{G_x}{F_0 \psi''} \ ,
\ee
where the right hand side is of the form $f (x,t) \phi (\dot x)$ for some functions $f, \phi$. Indeed, it is enough to take
 $G_x = f F_0$ oraz $\psi'' = 1/\phi$.

\ods
\begin{prop}  \label{prop-psi}
The equation $\ddot x = f (x,t) R (\dot x) $ admits a Lagrangain description with
the Lagrangian $L = \Psi (\dot x) + G (x, t)$, where
\be
   \Psi (v) := \int^v d\eta \int^\eta \frac{d \xi}{R (\xi) } \ , \quad G (x,t) = \int^x f (\xi, t) d \xi \ ,
\ee
(provided that the above integrals exist).
\end{prop}

\ods 

\begin{cor}

Special cases of Proposition~\ref{prop-psi}:

\begin{enumerate}

\item \  ${\ddot x} = \dot x f(x, t)$ \quad $\Longrightarrow$   \quad
$\dis {\cal L} =  {\dot x} \ln|\dot x| \ + \int^x f(\xi ,t)d \xi $

\item  \ ${\ddot x} = {\dot x}^2 f(x, t)$ \quad $\Longrightarrow$   \quad
$\dis {\cal L} =  - \ln|\dot x| \ + \int^x f(\xi, t) d\xi  $

\item \ ${\ddot x} = - k_0 {\dot x}^\nu $ \quad  $\Longrightarrow$   \quad
$\dis {\cal L} = \frac{ {\dot x}^{2-\nu}}{(2-\nu)(1-\nu)} - k_0 x $ \quad
($\nu \neq 1, 2$)

\item \ $\dis {\ddot x} = f (x,t) \left( 1 - \frac{{\dot x}^2}{c^2} \right)^{3/2} $
\quad $\Longrightarrow$   \quad
$\dis {\cal L} = - c^2 \sqrt{1 - \frac{{\dot x}^2}{c^2} } + \int^x f(\xi, t) d\xi  $

\end{enumerate}

\end{cor}

\ods

\section{Radical Lagrangians}

We consider Lagrangians of the form 
\be\label{1} 
{\cal L}
=\sqrt[\mu]{A (x,t) {\dot x}^{\nu} + B (x, t)}. 
\ee 
The Euler-Lagrange equations yield
\be  \label{rr-rad} 
{\ddot x}=\frac{p\dot x^{2\nu}+ q\dot
x^{2\nu-1}+ r \dot x^\nu+ s \dot x^{\nu-1}+ w}{g \dot x^{2\nu-2} +
h\dot x^{\nu-2}}, 
\ee 
where 
\be  \ba{l}  \dis g
:=\frac{(\nu-\mu)}{(1-\mu)}A \ , \qquad h
:=\frac{\mu(\nu-1)}{(1-\mu)}B \ ,  \\[4ex]\dis p:=-
\frac{(\nu+\mu)}{\nu(1-\mu)}A_x \ , \quad q:=-\frac{1}{(1-\mu)}A_t
\ ,  \\[4ex]\dis
r:=-\frac{(\nu-\nu\mu-\mu)}{\nu(1-\mu)}B_x-\frac{\mu(\nu+1)}{\nu(1-\mu)}\frac{A_x
B}{A} \ ,
\\[4ex]
\dis s:= - B_t-\frac{\mu}{(1-\mu)}\frac{A_tB}{A} \ , \quad
w:=\frac{\mu}{\nu(1-\mu)}\frac{B_x B}{A} \ . 
\ea \ee

\ods 

In this paper we will assume either $\mu = \nu \neq 1$ or $\mu \neq \nu = 1$. In those cases the denomiator simplifies and the righ-hand side of \rf{rr-rad} is a polynomial in $\dot x$. 

\subsection{The case $\mu = \nu \neq 1$ }

In this case the equation \rf{rr-rad} reduces to 
\be \label{5} 
\ddot x = \frac{\frac{2 A_x}{B } {\dot
x}^{\nu + 2} + \frac{A_t}{B} {\dot x}^{\nu + 1} + \left(
\frac{(1+\nu) A_x}{A} - \frac{\nu B_x}{B} \right) {\dot x}^2 +
\left( \frac{(1-\nu) B_t}{B} + \frac{\nu A_t}{A} \right) {\dot x}
+ \frac{B_x}{A} {\dot x}^{2-\nu} }{\nu (1-\nu)}.
\ee 
A further reduction is obtained by assuming that 
$A=A(t)$, $B=B(t)$.  Then the equation \rf{5} becomes
\be  \label{ABnu}
\ddot x = - \left( \frac{\dot A}{(\nu - 1) A} - \frac{\dot
B}{\nu B}  \right) \dot x -  \frac{\dot A}{\nu (\nu - 1) B} {\dot
x}^{\nu+1}.
\ee

\ods
\begin{prop}\label{prop-nunu}
The equation
\be  \label{abnu}
\ddot x = - a (t) \dot x - b (t) {\dot x}^{\nu + 1} 
\ee 
admits
(for $\nu \neq 0$, $\nu \neq 1$  and any functions $a, b$)  a
Lagrangian description with the Lagrangian of the form ${\cal L}
=\sqrt[\nu]{A (t) {\dot x}^{\nu} + B (t)}$ where \be \ba{l} A (t)
= \left( \nu \int^t b (\tau)  \exp \left( - \nu \int^\tau a (y) d
y  \right) d \tau \right)^{1-\nu}  , \\[3ex] B (t) = \left( \nu
\int^t b (\tau) \exp \left( - \nu \int^\tau a (y) d y \right) d
\tau \right)^{- \nu} \exp \left( - \nu \int^t a (\tau) d \tau
\right)  . \ea \ee
\end{prop}

In order to proof this proposition it is enough to compare \rf{abnu} with \rf{ABnu} and to solve resulting differential equations.  
\ods

Two interesting special cases can be obtained by requiring either $b = 0$ (i.e., $A (t) = \const$) or $a = 0$ (i.e.,  $\nu \ln A - (\nu - 1) \ln B = \const$).  

\ods

\begin{cor}  \label{cor-nunu}

Special cases of Proposition~\ref{prop-nunu}:

\begin{enumerate}

\item $A= A_0 = \const$, $B=B(t)$:
\[
\ddot x + a(t)\dot x= 0 \quad \Longrightarrow \quad 
{\cal L} = \sqrt[ \nu]{A_0 {\dot x}^\nu + B_0 \exp (-\nu\int^t b(\tau)d\tau)}
\] 
where $B_0 = \const$ and $\nu \neq 0, 1$.

\item $B = c_0 A^{\frac{\nu}{\nu - 1}}$, 
\[
  \ddot x + b (t) {\dot x}^m = 0 \quad \Longrightarrow \quad 
{\cal L} = \sqrt[m+1]{ F^{-m} {\dot x}^{m+1} + c_0 F^{-m-1} } 
\]
where $m \neq 1, 2$, $F = F (t) = - c_0 (m + 1) \int^t b (\tau) d \tau$, and $c_0 = \const$.

\end{enumerate}
\end{cor}
\ods

\subsection{The case $\mu \neq \nu = 1$ }

In this case 
 the equation \rf{rr-rad} reduces to
\be\label{6} 
{\ddot x}=\frac{A_x (1+\mu)\dot
x^{2}+\left(A_t+B_x(1-2\mu)+\frac{2\mu A_xB}{A}\right)\dot
x + B_t(1-\mu)+\frac{\mu A_t B}{A} - \frac{\mu B_x B }{A} }{A(\mu-1)}.
\ee 
We assume $A = A(t)$, $B= B (t)$. Then the equation \rf{6} becomes 
\be {\ddot
x}=\frac{\dot A}{(\mu-1)A}\dot x+\frac{\mu \dot A
B}{(\mu-1)A^2}-\frac{\dot B}{A} \ , 
\ee
and, solving linear differential equations (similarly as in the case of Proposition~\ref{prop-nunu}), we get
the following result. 
\ods

\begin{prop}\label{prop-ab2}
The equation
\be
\ddot x = a (t) \dot x + b (t) 
\ee 
admits (for any functions $a, b$)  a Lagrangian
description with the Lagrangian of the form ${\cal L}
=\sqrt[\mu]{A (t) {\dot x} + B (t)}$ where $\mu \neq 1$ 
 and 
\be \ba{l}  \label{AB} 
A (t) = \exp
\left((\mu-1)\int^t a (\tau)d \tau\right)  ,
\\[3ex] 
B (t) = -  \left( \int^t b (\tau) e^{- \int^\tau a (y) d y } d \tau \right) \exp
\left(\mu \int^t a (\tau)d \tau \right) . 
\ea \ee
\end{prop}

We point out that for $b = 0$ formulas \rf{AB} yield   
$B (t) = \exp
\left( \mu \int^t a (\tau)d \tau \right)$ (the integration constant has to be taken into account). 

\section{Multi-Lagrangian cases}  \label{sec-multi}

The Lagrangian of Proposition~\ref{prop-ab2} can be rewritten as
\be
   {\cal L} = e^{\int^t a (\tau) d\tau} \ \sqrt[\mu]{ {\dot x} e^{- \int^t a (\tau) d\tau} - \int^t b (\tau) e^{ - \int^\tau a (y) d y} } \ , 
\ee
and this form suggests the following generalization which can be easily verified by a simple straightforward calculation.  
 
\ods
\begin{prop}  \label{prop-F}
The equation
\be
\ddot x = a (t) \dot x + b (t) 
\ee 
admits (for any functions $a, b$)  a Lagrangian
description with the Lagrangian of the form 
\be
{\cal L} = e^{\int^t a (\tau) d\tau} \  F \left( {\dot x} e^{- \int^t a (\tau) d\tau} - \int^t b (\tau) e^{ - \int^\tau a (y) d y} \right)  \ ,
\ee
where $F$ is a function of one variable (such that $F '' \neq 0$).  
\end{prop} 

In particular, we point out that the simple classical equation $\ddot x + k \dot x = 0$ has Lagrangians of all forms considered in our paper, namely:
\be  \ba{l}  \label{multiL} \dis
 L_1 = \frac{1}{2} e^{k t} {\dot x}^2 \ , \quad \dis
L_2 = \frac{1}{e^{2 kt} \dot x  + e^{kt} } \ , \quad \dis
L_3 = {\dot x}^\mu e^{(\mu - 1) kt } \ , \\[3ex] \dis
L_4 = {\dot x} \ln |\dot x| - k  x \ , \quad 
L_5 = \sqrt[\nu]{{\dot x}^\nu + e^{ - \nu k t} } \ . 
\ea \ee
However, most of these forms (except $L_4$) can be reduced to the Lagrangian of Proposition~\ref{prop-F}, namely  
\be
L_F = e^{- k t} F ( {\dot x} e^{ k t} + c_0 ) \ , 
\ee
with $F (\xi)$ equal to $\frac{1}{2} \xi^2$, $\xi^{-1}$, $\xi^\mu$, $\sqrt[\nu]{\xi}$, respectively.  
\ods

Another multi-Lagrangian case is described by Corollary~\ref{cor-n-param}, where we present  
 a one-parameter family of Lagrangians for the equation $\ddot x + k {\dot x}^2 = 0$. What is more, the corresponding Hamiltonian is proportional to the Lagrangian (for any $n \neq 0$) and is time-independent. Hence, ${\cal L}$ is an integral of motion. 
This observation can be generalized as follows.

\ods

\begin{prop}  \label{prop-FL}
Suppose that a Lagrangian ${\cal L} = {\cal L} (q^i, {\dot q}^i, t)$ is an invariant of motion (i.e., $d{\cal L}/d t = 0$).
 Then, for any (sufficiently smooth) function $F : \R \rightarrow \R$, the Lagrangian ${\tilde {\cal L}} = F (\cal L)$ yields the same equations of motion.
\end{prop}

\no The proof is straightforward. We compute:
\[
  \frac{\partial \tilde {\cal L} }{\partial q^i} =  \frac{  d F }{d {\cal L} }  \frac{\partial {\cal L} }{\partial q^i} \ , \quad
  \frac{\partial \tilde {\cal L} }{\partial {\dot q}^i} = \frac{  d F }{d {\cal L} }  \frac{\partial {\cal L} }{\partial {\dot q}^i} \ , \quad
\frac{d }{d t} \frac{\partial \tilde {\cal L} }{\partial {\dot q}^i} = \frac{  d^2 F }{d {\cal L}^2 }  \frac{ d  {\cal L} }{ d t } +
\frac{ d F }{d {\cal L} }  \frac{d}{d t} \frac{\partial {\cal L} }{\partial {\dot q}^i} \ .
\]
Therefore,
\[
\frac{d }{d t} \frac{\partial \tilde {\cal L} }{\partial {\dot q}^i} -
  \frac{\partial \tilde {\cal L} }{\partial q^i} = \left(
\frac{d}{d t} \frac{\partial {\cal L} }{\partial {\dot q}^i} - \frac{\partial {\cal L} }{\partial q^i} \right) \frac{  d F }{d {\cal L} }  + \frac{  d^2 F }{d {\cal L}^2 }  \frac{ d  {\cal L} }{ d t } \ ,
\]
from which the proof follows immediately.
\ods

Taking into account Proposition~\ref{prop-FL} we see that ${\cal L}_F := F (\dot x e^{k t})$
is a Lagrangian for the equation $\ddot x + k {\dot x}^2 = 0$ (for any smooth function $F$).
Another Lagrangian (time-independent) for this equation was found by Sarlet:
${\cal L} = {\dot x} (1 - \ln {\dot x}) \exp (k x)$, see \cite{Sar1}.

\section{Conclusions}

In this paper we succeded to rederive all results of \cite{MRS,Mus}
in a straightforward, simple way. Actually, we found many other dissipative-looking systems possessing a Lagrangian description. 
One-dimensional systems admitting the Lagrangian formulation were discussed in numerous papers (see, e.g., \cite{KRS,CRS,CSL,Sar1,Kob,Leu,LH,Lo1,Gon,LL,LLG}), some of them devoted mostly to the damped harmonic oscillator, e.g., \cite{Lem1,Lem2,Dek}.
Surprisingly enough, using quite elementary tools, we succeeded to find some number of Lagrangians which seem to be overlooked in the exisiting literature.

It is interesting, that for some systems the Lagrangian description is not unique: they may belong to several classes (the problem of the equivalence was first discussed in \cite{CS}). 
The equations $\ddot x + k \dot x = 0$ and $\ddot x + k {\dot x}^2 = 0$, usually considered as classical dissipative equations (compare  \cite{Bau}),  have  infinite families of Lagrangians, see Section~\ref{sec-multi}.    The equation $\ddot x + k \dot x = 0$ has Lagrangians of all forms considered in our paper, see \rf{multiL}.

\end{document}